\newif\iflong
\newif\ifshort
\newif\ifjournal

\iflong
\else
\shorttrue
\fi

\documentclass[runningheads]{llncs}
\usepackage[T1]{fontenc}
\usepackage{graphicx}

\usepackage{xspace}

\usepackage{amsmath}
\usepackage{amssymb}
\usepackage{mathtools}
\usepackage{nicefrac}

\usepackage{hyperref}
\hypersetup{
    pdfencoding=auto, 
    psdextra,
    colorlinks=true,
    citecolor=green!40!black,
    linkcolor=red!50!black,
    urlcolor=blue!80!black
}

\usepackage{cleveref}

\usepackage[inline]{enumitem}
\usepackage[normalem]{ulem} %

\usepackage{caption}
\usepackage{subcaption}

\usepackage{pifont}
\newcommand{\cmark}{\ding{51}}%
\usepackage[fixed]{fontawesome5}

\usepackage{comment}

\usepackage{tabularx}
\usepackage{booktabs}
\usepackage{multirow}
\newcolumntype{Y}{>{\centering\arraybackslash}X}
\newcolumntype{M}[1]{>{\centering\arraybackslash}m{#1}}
\usepackage{colortbl}
\usepackage[most]{tcolorbox}

\usepackage{todonotes}

\usepackage{tikz}
\usetikzlibrary{decorations.pathreplacing}

\newcommand{\N}{\ensuremath{\mathbb{N}}}

\newcommand{\probName}[1]{\textsc{#1}\xspace}

\newcommand{\CCDC}[2][]{\probName{#2-Constructive Control by #1Deleting Projects}}
\newcommand{\CCDCs}[2][]{\probName{#2-CC#1DC}}
\newcommand{\DCDC}[2][]{\probName{#2-Destructive Control by #1Deleting Projects}}
\newcommand{\DCDCs}[2][]{\probName{#2-DC#1DC}}
\newcommand{\CCAC}[2][]{\probName{#2-Constructive Control by #1Adding Projects}}
\newcommand{\CCACs}[2][]{\probName{#2-CC#1AC}}
\newcommand{\DCAC}[2][]{\probName{#2-Destructive Control by #1Adding Projects}}
\newcommand{\DCACs}[2][]{\probName{#2-DC#1AC}}

\newcommand{\RXthreeC}{\probName{Restricted Exact Cover by~$3$-Sets}}
\newcommand{\RXthreeCs}{\probName{RX$3$C}}

\newcommand{\voters}{\ensuremath{V}} %
\newcommand{\numVoters}{\ensuremath{n}} %

\newcommand{\projects}{\ensuremath{P}} %
\newcommand{\numProjects}{\ensuremath{m}} %
\newcommand{\numCandidates}{\ensuremath{m}} %

\newcommand{\budget}{\ensuremath{B}} %
\newcommand{\cost}{\ensuremath{\operatorname{cost}}} %
\newcommand{\numCosts}{\ensuremath{c}} %
\newcommand{\sat}{\ensuremath{\operatorname{sat}}} %
\newcommand{\numDeleted}{\ensuremath{r}} %

\newcommand{\score}[1]{\ensuremath{\operatorname{score}_{\text{#1}}}}

\newcommand{\ilp}{\textsc{ilp}\xspace}
\newcommand{\del}{\textsc{del}}
\newcommand{\exh}{\textsc{exh}}

\newcommand{\ruleName}[1]{\textsc{#1}\xspace}
\newcommand{\greedyAV}{\ruleName{GreedyAV}}

\newcommand{\greedyCost}{\ruleName{GreedyCost}}

\renewcommand{\P}{\textsf{P}\xspace}
\newcommand{\NP}{\textsf{NP}\xspace}
\newcommand{\NPh}{\NP-hard\xspace}
\newcommand{\NPhness}{\NP-hardness\xspace}

\newcommand{\W}[1][1]{\textsf{W[#1]}}
\newcommand{\Wh}[1][1]{\W[#1]-hard\xspace}
\newcommand{\Whness}[1][1]{\Wh{}ness\xspace}
\newcommand{\Wc}[1][1]{\W[#1]-complete\xspace}
\newcommand{\Wclass}[1][1]{\textsf{W}\xspace}
\newcommand{\FPT}{\textsf{FPT}\xspace}
\newcommand{\XP}{\textsf{XP}\xspace}

\newcommand{\Oh}[1]{{\mathcal{O}(#1)}}

\spnewtheorem{observation}{Observation}{\bfseries}{\itshape}
\Crefname{observation}{Observation}{Observations}

\spnewtheorem{claim}{Claim}{\bfseries}{\itshape}
\Crefname{claim}{Claim}{Claims}
\let\doendproof\endproof
\renewcommand\endproof{~\hfill$\square$\doendproof}

\renewenvironment{example}{\begin{tcolorbox}[enhanced,breakable,colback=gray!5,colframe=gray!5,sharp corners,left=4pt,right=4pt,bottom=4pt,top=4pt]\begin{myExample}}{\end{myExample}\end{tcolorbox}}
\spnewtheorem{myExample}{Example}{\bfseries}{\itshape}
\Crefname{myExample}{Example}{Examples}

\usepackage{pict2e}
\newcommand{\opentriangle}{%
  \raisebox{0.2pt}{\makebox[0.77778em]{%
    \setlength{\unitlength}{0.6em}%
    \linethickness{0.4pt}%
    \begin{picture}(1,1)
    \polygon(0,0)(1,0)(1,1)
    \end{picture}%
  }}%
}
\newenvironment{proofsketch}[1]{\emph{Proof sketch.}\hspace{0.15cm}#1}{\hfill$\opentriangle$\medskip}

\newcommand{\Yes}{\textsc{Yes}\xspace}
\newcommand{\No}{\textsc{No}\xspace}

\author{%
    Šimon Schierrreich\inst{1,2} \and
    Krzysztof Sornat\inst{1}
}
\authorrunning{Šimon Schierreich and Krzysztof Sornat}
\institute{%
    AGH University of Krakow, Poland \and
    Czech Technical University in Prague, Czechia\\
    \email{schiesim@fit.cvut.cz}, \email{sornat@agh.edu.pl}%
}

\begin{document}
\title{Algorithms for Candidate Control in Greedy Participatory Budgeting Rules}
\maketitle              %

\begin{abstract}
    We study the problem of candidate control in participatory budgeting elections. Our focus is on two prominent sequential welfare-based rules---\greedyAV and \greedyCost---which are widely used in practice. Candidate control asks whether we can strategically modify the set of available candidates so as to either ensure that a preferred candidate $p$ is selected %
    or prevent~$p$ from being selected.%

    Since all variants of candidate control under the two rules we consider are known to be \NPh, we analyze the problems through the lens of parameterized complexity and approximability. Under the first lens, we provide a comprehensive classification with respect to natural parameters such as the number of voters, the number of controlled candidates, and the number of distinct costs, as well as their combinations. Within the second perspective, we establish a tight approximability bound.

\end{abstract}

\section{Introduction}\label{sec:intro}

Together with \emph{manipulation} and \emph{bribery}~\cite{FaliszewskiHH2009}, \emph{control} is a classic example of strategic interference in elections, and the study of its computational aspects dates back to the pioneering work of Bartholdi III \emph{et al.}~\cite{BartholdiTT1992}. The question in control is whether we can change the outcome of an election by changing its structure~\cite{BartholdiTT1992,FaliszewskiR2016}. More specifically, the prevailing modifications of the election's structure that are studied in the literature include the addition or deletion of candidates or voters. Traditionally, the goal of such attacks is either \emph{constructive}~\cite{BartholdiTT1992}---to make a distinguished candidate win---or \emph{destructive}~\cite{HemaspaandraHR2007}---to make some initially winning candidate lose.

Although most studies on control focus on single- and multi-winner voting~(see, e.g., the recent survey of Chen \emph{et al.}~\cite{ChenKNRSS2025}), we extend this line of research to \emph{participatory budgeting} (PB). PB is a modern democratic innovation, in which a municipality allows its citizens to decide how a certain fraction of its budget will be spent~\cite{Cabannes2004,AzizS2021,ReySM2025}. Participatory budgeting processes usually consist of two phases. In the first phase, significantly less explored in the computer science literature (the only exception we are aware of is the work of Rey \emph{et al.}~\cite{ReyEH2021}), citizens are invited to propose projects they would like to see implemented in their community. In the second phase, all inhabitants vote on the proposed projects.

Citizens commonly vote by casting their \emph{approval ballots}, meaning that each voter submits a set of projects they approve. These ballots are then aggregated using a voting rule. The most widely used rule is the so-called \greedyAV rule~\cite{TalmonF2019}, which iteratively selects the project with the highest support among the remaining projects that can still be funded within a budget limit. Similarly, under the \greedyCost rule~\cite{TalmonF2019}, the projects are considered in non-increasing order of their support to cost ratio. %
Both of these rules greedily maximize certain notions of social welfare---in the case of \greedyAV, measured as the sum of scores of selected projects, and in the case of \greedyCost, measured as the total voter support weighted by cost efficiency. Unlike the corresponding exact welfare-maximizing rules, both \greedyAV and \greedyCost are computable in polynomial time~\cite{TalmonF2019}.

The computational study of bribery and control in participatory budgeting has emerged only recently. Boehmer \emph{et al.}~\cite{BoehmerFJPPSSS2024} initiated this direction by introducing several bribery-like problems, motivated by the goal of making PB outcomes more explainable. Because most PB rules produce dichotomous results---each project is either funded or not---it can be difficult for organizers to explain to voters and to project proposers why a project succeeded or failed, particularly under more complex rules. Similar questions were addressed by Baumeister and Hogrebe~\cite{BaumeisterH2023}, and by Boehmer \emph{et al.}~\cite{BoehmerFJK2023}, who analyzed the robustness of PB outcomes via the \textsc{Flip-Bribery} problem~\cite{FaliszewskiHH2009}.

This line of research was further developed by Faliszewski \emph{et al.}~\cite{FaliszewskiJKPSSS2025a}, who introduced several performance measures based on \emph{candidate control} and conducted an extensive experimental analysis on real-world PB instances from PabuLib~\cite{FaliszewskiFPPSSST2023}. An example of such a measure is the number of projects one needs to remove from the instance to make some initially losing project win; the lower this number is, the closer the project is to winning. They demonstrated that these measures shed light on the internal structure of real-world PB instances, explaining how projects interact and compete within a given instance. However, they also proved that computing successful control actions in PB elections is \NPh for all combinations of control goals and operations, even under simple rules such as \greedyAV and \greedyCost. This result highlights the need for algorithms that go beyond brute-force enumeration and scale more effectively with the size of real-world PB elections.%

\begin{figure}
    \centering
    \definecolor{myGreen}{HTML}{A2F49B}
    \definecolor{myRed}{HTML}{E40515}
    \definecolor{myOrange}{HTML}{FB9A29}
    \definecolor{myYellow}{HTML}{FEE391}
    \begin{tikzpicture}[node distance=1.5cm and 1.5cm]
        \tikzstyle{every node} = [minimum height=0.6cm,,minimum width=0.8cm,fill=gray!10]
		\tikzstyle{result} = [draw,thick]
        \tikzstyle{partialresult} = [draw,dashed,thick]
		\tikzstyle{known} = [label={[label distance=-5pt]90:\small{\tt Known}}]
		\tikzstyle{NPh} = [fill=myRed!60]
		\tikzstyle{Wh} = [fill=myOrange!60]
		\tikzstyle{XP} = [fill=myYellow!60]
		\tikzstyle{FPT} = [fill=myGreen]

		\node[FPT,result] (m) at (0, 0) {$\numCandidates$};
        \node[NPh,known] (n) [right of=m] {$\numVoters$};
        \node[XP,Wh,result] (r) [right of=n] {$\numDeleted$};
        \node[XP,Wh,result] (q) [right of=r] {$q$};
        \node[XP,result] (c) [right of=q] {$\numCosts$};

        \node[XP,Wh,result] (nr) [below of=m,xshift=0cm,yshift=-0.3cm] {$\numVoters + \numDeleted$};
        \node[XP,Wh,result] (nq) [right of=nr] {$\numVoters + q$};
        \node[FPT,partialresult] (nc) [right of=nq] {$\numVoters + \numCosts$};
        \node[FPT,result] (rc) [right of=nc] {$\numDeleted + \numCosts$};
        \node[FPT,result] (qc) [right of=rc] {$q + \numCosts$};

        \node[FPT,result] (nrc) [below of=nq,xshift=0.5cm] {$\numVoters + \numDeleted + \numCosts$};
        \node[FPT,result] (nqc) [right of=nrc,xshift=0.5cm] {$\numVoters + q + \numCosts$};

        \draw[<->] (r.south west) -- (nr.north east);
        \draw[<->] (q.south west) -- (nq.north east);
        \draw[->] (rc.south) -- (nrc.north);
        \draw[->] (qc.north) -- (c.south);
        \draw[->] (qc.south) -- (nqc.north);
    \end{tikzpicture}
    \caption{A summary of parameterized complexity results for \greedyAV and \greedyCost in both constructive and destructive settings. Each box represents a parameter combination: $\numCandidates$ (number of candidates), $\numVoters$ (number of voters), $\numDeleted$ (number of controlled candidates), $\numCosts$ (number of distinct costs). Colors indicate complexity: green = \FPT, yellow = \XP, orange = \Wh $\cap$ \XP, red = \NPh for constant parameter values. An arrow from $A$ to $B$ indicates that the result for $A$ implies the same for $B$. Bordered boxes mark results proven in this work; dashed borders indicate results that additionally require a specific (but natural) tie-breaking rule.}
    \label{fig:AV:paramComplexityResults}\label{fig:AVcost:paramComplexityResults}
\end{figure}

\subsection{Our Contribution}

In this paper, we focus on the algorithmic aspects of candidate control in participatory budgeting under two prominent PB rules---\greedyAV and \greedyCost. Our main motivation is to develop efficient algorithms for the control problems that underlie the performance measures proposed by Faliszewski \emph{et al.}~\cite{FaliszewskiJKPSSS2025a}. In view of the hardness results established there, we employ parameterized complexity and approximation algorithms frameworks, which equip us with formal tools for a more fine-grained analysis of the computational complexity of (hard) computational problems.

From the perspective of parameterized complexity, our goal is to identify the sources of computational intractability and to design algorithms that become efficient when certain aspects of the input are restricted. Common restrictions in voting theory include a bounded number of projects (denoted by $\numProjects$) or voters (denoted by $\numVoters$) ~\cite{ConitzerSL2007,ChenFNT2017}. Beyond these standard parameters, we also consider the number of projects~$\numDeleted$ that can be controlled, the number of non-controlled projects $q$, and the number of distinct project costs $\numCosts$. 
We would like to highlight here that both parameters, $\numDeleted$ and $\numCosts$, are expected to be small in practice. In the former case, Faliszewski \emph{et al.}~\cite{FaliszewskiJKPSSS2025a} showed that for most projects in real-life PB instances, $\numDeleted$ rarely exceeds the value of $10$. In the latter case, costs are often multiples of a fixed unit (e.g., 100\,000~EUR) or proposers simply declare the maximum cost in a project category, leading to few distinct values.
For these parameters and their combinations, we provide classification of the corresponding problems with respect to the complexity classes \FPT, \XP, and \Wh.{} See \Cref{fig:AV:paramComplexityResults} for an overview of our results.

The second perspective we consider is approximability. Here, instead of restricting the input, we seek polynomial-time algorithms that work for all instances, at the cost of allowing solutions whose values approximate the optimum (since the problems are NP-hard).
Unfortunately, our results are largely negative, establishing strong inapproximability. In particular, no $m^{1-\epsilon}$-approximation is possible for any $\epsilon > 0$ (unless $\P = \NP$), yet a simple $m/c$-approximation algorithm exists (for every fixed $c \geq 1$). Hence, the obtained inapproximability ratio is high but essentially tight.

Together, these results provide the first fine-grained computational complexity study of hard control problems in participatory budgeting, offering a detailed understanding of when and how candidate control in PB can be exactly computed or approximated. The most important application of our work is that, in order to compute the performance measures of Faliszewski \emph{et al.}~\cite{FaliszewskiJKPSSS2025a} in practice, one should rely on parameterized algorithms rather than approximation algorithms. Indeed, Faliszewski \emph{et al.}~\cite{FaliszewskiJKPSSS2025a} showed that $r$ tends to be small in real-world settings, so we believe that our algorithmic results are a necessary step toward widespread adoption of their measures.

\subsection{Related Work}

From the perspective of classical complexity, deciding whether a successful attack on an election---through manipulation, bribery, or control---has been shown to be computationally intractable for many popular voting rules. This holds across single-winner~\cite{BartholdiTT1989,BartholdiTT1992,FaliszewskiHH2009,FaliszewskiR2016,ConitzerW2016,ErdelyiNRRYZ2021}, multi-winner~\cite{MeirPRZ2008,CaragiannisKM10,AzizGGMMW2015,FaliszewskiST2017}, and participatory budgeting~\cite{BoehmerFJPPSSS2024,FaliszewskiJKPSSS2025a} settings; see also the survey of Chen \emph{et al.}~\cite{ChenKNRSS2025}.%

Following these intractability results, many authors have studied strategic behavior in elections through the lens of parameterized complexity~{}\ifshort{}\cite{CyganFKLMPPS2015}\fi{}. The most prominent parameters considered are the number of candidates~\cite{ConitzerSL2007,Yang2014} and the number of voters~\cite{ChenFNT2017}. Other parameters include the number of affected voters, the available budget for an attack, structural properties of the election, and the number of selected candidates in multi-winner settings~\cite{BredereckCFNN16,YangG2015,FaliszewskiST2017,Yang2019,Yang2020,KusekBFKK23}. A survey of parameterized complexity results for single-winner elections is provided by Betzler \emph{et al.}~\cite{BetzlerBCN2012}. The only works studying the parameterized complexity of strategic interference in PB are those of Boehmer \emph{et al.}~\cite{BoehmerFJPPSSS2024,BoehmerFJK2023}, who present several parameterized results for bribery-like problems. Importantly, we are not aware of any papers addressing the parameterized complexity of control in PB.

Beyond fixed-parameter analysis, another way to address \NPhness is through approximation algorithms. This approach to attacks on elections has received significantly less attention than the parameterized-complexity viewpoint. The few existing studies focus mainly on bribery and manipulation in single-winner elections, analyzing approximation algorithms and hardness results for various rules~\cite{BrelsfordFHSS2008,Faliszewski2008,ZuckermanPR2009,ElkindF2010,KellerHH2019,FaliszewskiMS2021}. To the best of our knowledge, the only studies addressing control problems from an approximation perspective are those of Brelsford~\cite{Brelsford2007} and Faliszewski \emph{et al.}~\cite{FaliszewskiHH2015}. They showed that, under standard complexity assumptions, no algorithm can achieve an approximation ratio better than $\Oh{\log \numCandidates}$ for the corresponding single-winner rules. None of these works extend beyond the single-winner framework, and therefore their results do not generalize to participatory budgeting or other multi-winner environments. Indeed, Faliszewski \emph{et al.}~\cite{FaliszewskiJKPSSS2025a} showed that, under our PB rules, determining whether candidate control is possible is polynomial-time solvable in the multi-winner regime.

\section{Preliminaries}\label{sec:preliminaries}

We use $\N$ to denote the set $\{1,2,3,\ldots\}$ of natural numbers and we set $\N_0 = \N \cup \{0\}$. For $i\in\N$, we set $[i] = \{1,\ldots,i\}$ and $[i]_0 = \{0\} \cup [i]$. Furthermore, for a pair of integers $i,j\in \N$ such that $i < j$, we set $[i,j] = \{i,i+1,\ldots,j\}$ and call it an \emph{interval}. For a set $S$, we use $2^S$ to denote all subsets of $S$, including the empty set $\emptyset$ and~$S$ itself, and for an integer $k\in\N$, we denote by $\binom{S}{k}$ the set of all $k$-sized subsets of $S$. 

\paragraph{Participatory Budgeting.}

An instance of \emph{participatory budgeting} (PB) election is a quadruple $E=(\projects,\cost,\voters,\budget)$, where $\projects$ is a set of $\numProjects$ \emph{projects}, $\cost\colon\projects\to\N$ is a function assigning to each project $p\in\projects$ its \emph{cost} for which $p$ can be implemented, $\voters$ is a set of $\numVoters$ \emph{voters}, and $\budget\in\N$ is the available \emph{budget}. Each voter $v\in\voters$ casts its preferences using an \emph{approval ballot} $A(v) \subseteq \projects$, which is simply a set of projects voter $v$ approves. For a candidate $p\in\projects$, we denote by $A(p)$ the set of voters who approve the project $p$; that is, $A(p) = \{ v\in \voters \mid p \in A(v) \}$. We use $\score{}(p) = |A(p)|$ to denote the \emph{score} of the project $p$, that is, the number of its supporters.
Furthermore, we extend the definition of the cost function from a single project to sets of projects by setting $\cost(S) = \sum_{p\in S} \cost(p)$ for every $S\subseteq\projects$. {}\ifshort{}A set of projects $S\subseteq \projects$ is \emph{$\budget$-feasible} if $\cost(S) \leq \budget$.\fi{}

A participatory budgeting \emph{rule} is a function $f\colon E\to 2^{\projects}$ that, given an election $E$, returns a $\budget$-feasible set of projects. Let $W=f(E)$ for some election~$E$ and PB rule~$f$. We say that $W$ is a set of \emph{selected} (or equivalently \emph{funded}) projects under $f$. The projects in $\projects\setminus W$ are called \emph{losing}.%

Throughout the paper, we focus on two concrete examples of a larger family of \emph{greedy} participatory budgeting rules, which were defined by Talmon and Faliszewski~\cite{TalmonF2019}.\footnote{We extend their definition of a satisfaction function to incorporate dependence on a cost function.}
\begin{definition}%
    A participatory budgeting rule is \emph{greedy} if it is based on a \emph{satisfaction function} $\sat \colon 2^P \times 2^P \times \N^P \rightarrow \N_0$ and
    for given an election $E=(\projects,\cost,\voters,\budget)$
    it proceeds in iterations as follows.
    The rule starts with an empty set of funded projects $W = \emptyset$.
    In each iteration it adds a project $p$ to $W$ which maximizes the value $\sum_{v \in \voters} \sat(A(v), W \cup \{p\}, \cost)$ under a condition that $W \cup \{p\}$ is $\budget$-feasible.
    It ends when none of losing projects can be added to $W$ without violating $\budget$-feasibility.
\end{definition}

The first rule, \greedyAV uses satisfaction function $(A,B,\cdot) \mapsto |A \cap B|$, which does not depend on the cost function.
This implies that \greedyAV considers the projects in non-increasing order of their scores (with ties resolved by a predefined tie-breaking order), i.e.,
project $p$ is considered before $p'$ whenever $\score{}(p) > \score{}(p')$.
Second rule, \greedyCost uses satisfaction function $(A,B,\cost) \mapsto \frac{|A \cap B|}{\cost(A \cap B)}$.
This implies that \greedyCost considers the projects in non-increasing order of their score-to-cost ratios (with ties resolved by a predefined tie-breaking order), i.e.,
project $p$ is considered before $p'$ whenever $\frac{\score{}(p)}{\cost(p)} > \frac{\score{}(p')}{\cost(p')}$.

\newcommand{\pOne}{\ensuremath{p_1}\xspace}
\newcommand{\pTwo}{\ensuremath{p_2}\xspace}
\newcommand{\pThree}{\ensuremath{p_3}\xspace}
\newcommand{\pFour}{\ensuremath{p_4}\xspace}
\newcommand{\pFive}{\ensuremath{p_5}\xspace}
\newcommand{\pSix}{\ensuremath{p_6}\xspace}

\begin{figure}[bt!]
    \centering
    \renewcommand{\cmark}{\color{green!60!black}\ding{51}}
    \renewcommand{\arraystretch}{1.2}
    \begin{tabular}{M{1cm}|M{1cm}|M{1cm}|M{1cm}|M{1cm}|M{1cm}|M{1cm}}
        \toprule
              & \pOne & \pTwo & \pThree & \pFour & \pFive & \pSix  \\
              & \small{50} & \small{20} & \small{5} & \small{10} & \small{2} & \small{3} \\
              \midrule
        $v_1$ & \cmark & \cmark & \cmark &        &        & \cmark \\
        $v_2$ & \cmark &        & \cmark &        &        &        \\
        $v_3$ &        & \cmark &        & \cmark & \cmark &        \\
        $v_4$ & \cmark &        &        & \cmark &        & \cmark \\
        $v_5$ & \cmark & \cmark &        & \cmark &        &        \\\bottomrule
    \end{tabular}
    \caption{An example of a participatory budgeting election.}
    \label{fig:exampleInstance}
\end{figure}

\begin{example}\label{ex:instance}\itshape
    Let us assume the instance depicted in \Cref{fig:exampleInstance}.
    Furthermore, let the budget be $\budget=63$, and the tie-breaking order considers the projects lexicographically.
    \greedyAV considers the projects in the following order (under each project, we write its score).
    \[
        \pi = \left(
            \underset{4}{\pOne},
            \underset{3}{\pTwo},
            \underset{3}{\pFour},
            \underset{2}{\pThree},
            \underset{2}{\pSix},
            \underset{1}{\pFive} 
        \right)
    \]
    In the first round, the project \pOne is considered. Since its cost is $50$, which is smaller than $\budget$, we add it to $W$. In the second round, the rule considers \pTwo. However, $\cost(\pOne) + \cost(\pTwo) = 70 > \budget$, so \pTwo is skipped.
    The project \pFour, on the other hand, is affordable, so we add \pFour to $W$.
    The project \pThree is not affordable, so it is also skipped, while~\pSix can be funded. Therefore, a current solution is $W = \{\pOne,\pFour,\pSix\}$. Overall cost of this subset of projects is $63$, which is equal to the overall budget. Therefore, the remaining project \pFive cannot be funded and $W = \{\pOne,\pFour,\pSix\}$ is the outcome of the rule.

    \greedyCost considers the projects in the following order (each project is annotated with its score divided by its cost).
    \[
        \pi = \left(
            \underset{0.67}{\pSix},
            \underset{0.5}{\pFive},
            \underset{0.4}{\pThree},
            \underset{0.3}{\pFour},
            \underset{0.15}{\pTwo},
            \underset{0.08}{\pOne} 
        \right)
    \]
    With this specific ordering, the above process terminates with $W=\{\pSix,\pFive,\pThree,\pFour,\pTwo\}$. The total cost of $W$ is $40$.
\end{example}

\paragraph{Control Problems.}

Throughout the paper, we are interested in all possible combinations of two different \emph{goals}---\emph{constructive} and \emph{destructive}---and two different control \emph{operations}---project \emph{deletion} and project \emph{addition}. We follow the standard control notation for single- and multi-winner voting as given by Faliszewski and Rothe~\cite{FaliszewskiR2016}.

Let $f$ be a PB rule and \textsc{Op} be a control operation. If the goal is constructive, then the task is to perform a limited number of control operations \textsc{Op} to make an initially losing candidate win. On the other hand, if the goal is destructive, then we aim to make an initially winning candidate lose in the modified instance.

Under the project deletion operation, we can modify the instance by deleting a limited number of projects of $\projects\setminus\{p\}$, where $p$ is the distinguished project whose outcome we aim to influence. Formally, this operation is captured by the following computational problem.

\vspace{0.15cm}
\noindent\begin{tabularx}{\linewidth}{lX}
	\toprule
	\multicolumn{2}{c}{\CCDC{$f$}} \\\midrule
	\small\emph{Input:} & \small{} An election $E=(\projects,\cost,\voters,\budget)$, an integer $\numDeleted$, and a distinguished candidate $p\in\projects$. \\
	\small\emph{Question:} & \small{}Is there a set $X\subseteq \projects\setminus\{p\}$ of size at most $\numDeleted$ such that $p\in f((\projects\setminus X,\cost,\voters,\budget))$? \\
	\bottomrule
\end{tabularx}
\vspace{0.1cm}

\noindent{}We use \CCDCs{$f$} for short, following the standard terminology in election control (where projects correspond to candidates, so DC comes from \emph{Deleting Candidates}). The \DCDC{$f$} problem (\DCDCs{$f$}) is then defined analogously, we just require that $p\not\in f((\projects\setminus X,\cost,\voters,\budget))$. In both problems, we wlog assume $r \leq m-1$ because $r > m-1$ effectively means that we can delete all $m-1$ projects except~$p$.

\begin{example}\label{ex:destructiveControl}
    Consider the instance from \Cref{ex:instance} and recall that under \greedyAV rule, the set of funded projects is $W=\{\pOne,\pFour,\pSix\}$. Assume \CCDCs{\greedyAV} with $\numDeleted=1$ and $p=\pFive$. If we remove $X=\{\pSix\}$, then in round 5, \greedyAV considers $\pFive$ instead of $\pSix$ and we have $W_4 = \{\pOne,\pFour\}$ with a total cost of $60$. Therefore, $\pFive$ is affordable, and the rule funds it. Hence, we performed a successful constructive control by deleting projects. 
\end{example}

Under the project addition operation, we have two sets of projects: the set $\projects$ of \emph{standard projects} and additionally a set $Q$ of \emph{spoiler projects}. Slightly abusing the notation, we set $\numCandidates = |P| + |Q|$ if we are in the setting of adding candidates.
Rule $f$ does not initially consider spoiler projects, yet voters already have preferences over them. The question then is whether a subset of spoiler projects can be added to the instance to achieve a given control goal. Formally, this is captured by the following decision problem.

\vspace{0.15cm}
\noindent\begin{tabularx}{\linewidth}{lX}
	\toprule
	\multicolumn{2}{c}{\DCAC{$f$}} \\\midrule
	\small\emph{Input:} & \small{} An election $E=(\projects,\cost,\voters,\budget)$, a set of spoiler projects $Q$, an integer $\numDeleted$, and a distinguished candidate $p\in\projects$. \\
	\small\emph{Question:} & \small{}Is there a set $X\subseteq Q$ of size at most $\numDeleted$ such that $p\not\in f((\projects \cup X,\cost,\voters,\budget))$? \\
	\bottomrule
\end{tabularx}
\vspace{0.1cm}

\noindent{}We use \DCACs{$f$} for short, and the \CCAC{$f$} problem (\CCACs{$f$}) is defined analogously.

\paragraph{Parameterized Complexity.}

In parameterized complexity~\cite{DowneyF2013,Niedermeier2006,CyganFKLMPPS2015}, we analyze the running time of a computational problem with respect to the input size $|\mathcal{I}|$ and some additional parameter $k$. The most favorable outcome of such an analysis is an algorithm running in $f(k)\cdot |\mathcal{I}|^\Oh{1}$ time, where~$f$ is a computable function. Such algorithm is called \emph{fixed-parameter algorithm} and the complexity class containing all parameterized problems for which fixed-parameter algorithm is possible is called \FPT. The less favorable but still positive outcome is an algorithm running in $|\mathcal{I}|^{f(k)}$, where~$f$ is again a computable function. The complexity class containing all problems admitting such algorithms is called \XP.
Under the well established assumption that $\FPT\not=\W$, one can rule out the existence of a fixed-parameter algorithm for a parameterized problem by showing that this problem is \Wh. This can be done through a \emph{parameterized reduction} from some known \Wh problem. %

\section{Parameterized Complexity}

We start this section with a few observations that apply to any polynomial-time computable participatory budgeting rule and any of the studied variants of our control problems.

First, we observe that if the number of allowed changes is constant, we can solve all variants of control in polynomial time. The argument is based on an exhaustive enumeration of all subsets of projects we are allowed to control and simulation of the respective rule with these projects removed (added). An analogous approach applies to the dual parameter of the number of removed/added projects: the number~$q$ of projects that are not controlled.

\ifshort
\begin{observation}
\fi
    \label{thm:CCDC:XP:numDeleted}
    \label{thm:CCAC:XP:numDeleted}
    \label{thm:CCDC:XP:numNotDeleted}
    For every polynomial-time computable rule~$f$, \CCDCs{$f$}, \DCDCs{$f$}, \CCACs{$f$}, and \DCACs{$f$} are in~\XP when parameterized by the number of deleted/added projects~$\numDeleted$ or the number of unaffected projects~$q=\numProjects-\numDeleted$ ($q = |Q|-\numDeleted$ in case of project addition operation).
\end{observation}

Next, we note that parameterization by the number of projects~$\numCandidates$ yields a straightforward \FPT algorithm. This follows from the fact that we can check every subset of projects of size at most $\numDeleted$ to see whether removing (or adding) them achieves the control goal by applying the respective rule to the modified instance.

\begin{observation}\label{thm:AV:CCDC:FPT:projects}\label{thm:AV:DCDC:FPT:projects}
    For every PB rule~$f$ for which a winning project can be found in \FPT time when parameterized by the number of projects~$\numProjects$, \CCDCs{$f$}, \DCDCs{$f$}, \CCACs{$f$}, and \DCACs{$f$} are in~\FPT when parameterized by the number of projects~$\numProjects$.
\end{observation}

We start our exploration of parameterized complexity of candidate control by showing that the trivial algorithms of \Cref{thm:CCAC:XP:numDeleted} cannot be significantly improved; that is, the existence of fixed-parameter algorithms for any of these two parameters is unlikely.

\ifshort
\begin{theorem}[$\star$]
\else
\begin{theorem}
\fi
    \label{thm:AV:DCDC:Wh:voters:deletions}
    Both \DCDCs{\greedyAV} and \DCDCs{\greedyCost} are \Wh when parameterized by the number of deleted projects~$\numDeleted$ or the number of unaffected projects~$q = \numProjects-\numDeleted$, even if $n=2$.%
\end{theorem}
\ifshort
\begin{proofsketch}
    \newcommand{\base}{\ensuremath{b}}
	We show \Whness by a parameterized reduction from the \probName{Perfect Code} problem. The input of this problem consists of a simple undirected graph~$G=(U,E)$ and an integer~$k\in\N$. The goal is to decide whether there exists a set~$X\subseteq U$ of at most~$k$ vertices such that for every~$u\in U$, there is exactly one vertex in~$N_G[u]\cap X$. \probName{Perfect Code} is known to be \ifshort\Wh\fi{} when parameterized by~$k$, even if there is no solution~$X\subseteq U$ of size $k' < k$~\ifshort\cite{DowneyF1995}\fi.

    In the interest of space, we present the reduction only for the case of \DCDCs{\greedyAV}. Let~$\mathcal{I}$ be an instance of \probName{Perfect Code},~$(u_1,\ldots,u_{|U|})$ be an arbitrary but fixed ordering of its vertices, and let~$\base = \Delta(G) + 2$ be an integer. We construct an equivalent instance~$\mathcal{J}$ of \DCDCs{\greedyAV} as follows. The set of projects contains selection-projects~$p_1,\ldots,p_{|U|}$, which are in one-to-one correspondence with the vertices~$u_1,\ldots,u_{|U|}$ of~$G$, one guard-project~$g$, and the distinguished project~$p$. We use the costs of our selection-projects to encode the neighborhoods of the corresponding vertices. Formally, let~$p_i$,~$i\in[|U|]$, be a selection-project. Then, we set~$\cost(p_i) = \sum_{u_j\in N_G[u_i]} 1\cdot \base^j$. That is, the \mbox{base-$\base$} representation of agent~$p_j$'s cost contains one as its~$j$-th digit if and only if the vertex~$u_j$ is in the closed neighborhood of vertex~$u_i$ in~$G$. Otherwise, the digit is zero. To finalize the definition of projects' costs, we set~$\cost(p) = 1$,~$\cost(g) = (\sum_{i=1}^{|U|} 1\cdot \base^i) + 1$. Next, we add two voters,~$v_1$ and~$v_2$. Voter~$v_1$ approves all projects except for~$p$, and~$v_2$ approves only the selection-projects. Finally, we set~$\budget = (\sum_{i=1}^{|U|} \cost(p_i)) + 1$, and~$\numDeleted = k$. The general idea of the construction is that we need to remove selection-projects whose overall cost is exactly~$\sum_{i=1}^{|U|} 1\cdot \base^i$ or the project~$p$ is funded by \greedyAV. By the construction of selection-projects' costs, it follows that vertices corresponding to removed selection-projects form a perfect code.
\end{proofsketch}
\fi

The basic idea for the \Whness reduction for constructive control remains the same as in \Cref{thm:AV:DCDC:Wh:voters:deletions}. However, this time, we need to be more careful with the definition of our guard-projects. Due to a large number of these guards, this time we cannot easily tweak the reduction to also show \Whness for parameter~$q$.

\ifshort
\begin{theorem}[$\star$]
\fi
    \label{thm:AV:CCDC:Wh:voters:deletions}
    Both \CCDCs{\greedyAV} and \CCDCs{\greedyCost} are \Wh when parameterized by the number of deleted projects~$\numDeleted$, even if $\numVoters = 3$.
\end{theorem}

Note that the reductions in \Cref{thm:AV:DCDC:Wh:voters:deletions,thm:AV:CCDC:Wh:voters:deletions} use a constant number of voters and are polynomial. Thus, all project-deletion control problems are para-\NPh with respect to~$\numVoters$. This also follows from~\cite{FaliszewskiJKPSSS2025a}.%

In the remainder of this subsection, we present our tractability results for project deletion. The first algorithm for the combined parameter of the number of voters and the number of different costs is based on guessing, combined with an Integer Linear Programming (ILP) formulation of a carefully constructed sub-problem. A limitation of this approach is that the algorithm requires a specific (but very natural) tie-breaking rule in which all projects of the same cost (and with the same score) follow each other. An example of such tie-breaking is the \emph{cheaper-first} rule~\cite{BoehmerFJK2023}, which, in the event of a tie, always prefers a project of smaller cost (and resolves the tie between two projects of the same cost arbitrarily, e.g., lexicographically).

\begin{theorem}\label{thm:AV:CCDC:FPT:voters:costs}
    For any greedy PB rule $f$ which considers projects of equal score and cost consecutively, both \CCDCs{$f$} and \DCDCs{$f$} are \FPT when parameterized by~$\numCosts+\numVoters$, where~$\numCosts$ is the number of different costs and~$\numVoters$ is the number of voters.
\end{theorem}
\ifshort
\begin{proofsketch}
    Let $\mathcal{I}=(\projects,\voters,\budget,p)$ be an instance of the minimization variant of \CCDCs{\greedyAV} (i.e., we need to find the smallest set of projects whose deletion makes~$p$ funded). The optimal solution size can then be compared to~$r$ and the decision version of the problem solved immediately. %
    
    The \greedyAV algorithm processes the projects in non-increasing order according to their scores.
    Therefore, we can divide the order of projects into at most~$\numVoters+1$ \emph{blocks}, where projects from the same block have the same score.
    The cheaper-first tie-breaking rule in the case of \greedyAV implies that projects with the same score and the same cost are processed consecutively,
    and we call each of such a set of projects \emph{a sub-block}.
    Sub-blocks are disjoint.

    Without loss of generality, we can assume that~$p$ is the last project in the last sub-block.
    We have~$s \leq (\numVoters+1) \cdot \numCosts$ many non-empty sub-blocks, and we denote them by~$P_i \subseteq P$ for~$i \in \{1, 2, \dots, s\}$, such that \greedyAV processes sub-blocks in the order based on their indices.
    Therefore,%
    ~$p \in P_s$.

    In the following lemma, we show that a minimum-size feasible solution~$D$ to~$\mathcal{I}$ has a certain structure, which will be exploited by our algorithm.
    Namely, in every sub-block from which the solution~$D$ deletes a project, \greedyAV funds all non-deleted projects (due to enough remaining budget).
    Let~$B_i$ be the remaining budget \greedyAV has before processing projects from a sub-block~$P_i$ in an instance~$\mathcal{I}\setminus D$.
    
    \begin{lemma}[$\star$]\label{lem:structured-solution-n+c}\label{prop:sub-blocks-structure}
        For every~minimum-size feasible solution $D$ to~$\mathcal{I}$ and for every sub-block~$P_i$ it holds that:
        if~$P_i \cap D \neq \emptyset$ then~$B_i \geq \cost(P_i \setminus D)$. More precisely, exactly one of the following formulas holds:
        \begin{enumerate*}
            \item~$B_i \geq \cost(P_i \setminus D)$.
            \item~$P_i \cap D = \emptyset$ and~$B_i < \cost(P_i)$.
        \end{enumerate*}
    \end{lemma}

    Two exclusive formulas from \Cref{prop:sub-blocks-structure} implies two different behaviors of \greedyAV in a sub-block:
    the first means that \greedyAV funds all available projects from the sub-block;
    the second means that \greedyAV cannot buy all available projects from the sub-block.
    We can guess the behavior of \greedyAV in all sub-blocks by considering~$2^{s-1} \leq 2^{(\numVoters+1) \cdot \numCosts}$ cases
    (we only require that the last sub-block~$P_s$ satisfies formula 1 from \Cref{prop:sub-blocks-structure}---because in formula 2 project~$p$ would not be funded).
    For each guess we solve the ILP described below.
    Every solution to an ILP gives us a corresponding deletion set (some ILPs may not be feasible).
    A feasible solution with the smallest value of the objective function (among all the ILPs solved) corresponds to the smallest feasible solution~$D$.

    According to the guessed structure of the sub-blocks (due to \Cref{prop:sub-blocks-structure}) we partition the set of sub-blocks~$\{P_1, P_2, \dots, P_s\}$ into sets~$P_{\del}, P_{\exh}$, where
   ~$P_{\del}$ contains sub-blocks in which \greedyAV funds all non-deleted projects (deletions are possible; this is the formula 1 in \Cref{prop:sub-blocks-structure})
    and~$P_{\exh}$ contains sub-blocks in which \greedyAV exhausts the budget and does not fund all the projects (deletions are not allowed; this is the formula 2 in \Cref{prop:sub-blocks-structure}).
    Recall that we have~$P_s \in P_{\del}$.

    For every sub-block~$P_i \in P_{\del}$ we define a variable~$x_i$ which value means how many projects we delete from~$P_i$:
   ~$$x_i \in \N_0 \hspace{50pt} \forall P_i \in P_{\del},$$
    and we add a constraint meaning that we cannot delete more projects than the cardinality of~$P_i$:
   ~$$x_i \leq |P_i| \hspace{50pt} \forall P_i \in P_{\del}.$$
    For every sub-block~$P_i \in P_{\exh}$ we define a variable~$y_i$ which means how many projects from~$P_i$ \greedyAV funds:
   ~$$y_i \in \N_0 \hspace{50pt} \forall P_i \in P_{\exh},$$
    and we add a constraint meaning that \greedyAV cannot fund more projects than the cardinality of~$P_i$ minus~$1$ (because~$P_i \in P_{\exh}$):
    \begin{equation}
        y_i \leq |P_i|-1 \hspace{50pt} \forall P_i \in P_{\exh}.\label{eq:y-leq-size-Pi}
    \end{equation}
    The objective of the ILP is to minimize the number of deleted projects, i.e.,
   ~$$\min \:\:\: \sum_{P_i \in P_{\del}} x_i.$$
    Now, we model the execution of \greedyAV by adding the following variables and constraints.
    We define a variable~$B_i$ as the budget available to the algorithm before processing sub-block~$P_i$:
   ~$$B_i \in \N_0 \hspace{50pt} \forall P_i \in P.$$
    Naturally we fix~$B_1 = B$ and additionally we define a variable~$B_{s+1} \in \N_0$ which will ensure that we do not overspend the initial budget~$B$.
    We define a constant~$\cost_i$ as the cost of a project from~$P_i$.
    
    Now we can define the relation between~$B_i$'s that comes from the execution of \greedyAV.
    We have two types of constraints, depending on the membership of~$P_i$.
    
    If~$P_i \in P_{\del}$ then~$B_{i+1}$ is the remaining budget from~$B_i$ after paying for all non-deleted~$|P_i| - x_i$ projects:
    \begin{equation}
        B_{i+1} = B_i - (|P_i| - x_i) \cdot \cost_i \hspace{30pt} \forall P_i \in P_{\del}.\label{eq:bi-def-in-pd}    
    \end{equation}
    Note that the non-negativity of~$B_{i+1}$ can always be satisfied by simply setting~$x_i = |P_i|$.
    
    If~$P_i \in P_{\exh}$ then~$B_{i+1}$ is the remaining budget from~$B_i$ after paying for as many projects as possible from~$P_i$:
    \begin{align}
        B_{i+1} &= B_i - y_i \cdot \cost_i &\forall P_i \in P_{\exh},\label{eq:bi-def-in-pe}\\
        B_{i+1} &\leq \cost_i - 1 &\forall P_i \in P_{\exh}.\nonumber
    \end{align}
    The constraint \eqref{eq:y-leq-size-Pi} ensures that we cannot fund all projects from~$P_i \in P_{\exh}$.
    Note that these constraints may not be possible to satisfy, e.g., if~$B \geq \sum_{j \in [i]} \cost(P_j)$ then~$B_i \geq \cost(P_i)$ (so the guess~$P_i \in P_{\exh}$ was wrong).
    But if all the guesses are consistent with the execution of \greedyAV on an optimal solution, then~$y_i = \lfloor B_i / \cost_i \rfloor$ is feasible.

    On the other hand, every feasible solution~$(x^{\ilp}_i,y^{\ilp}_i,B^{\ilp}_i)_{i \in [s]}$ to an ILP with a guessed partition~$(P^{\ilp}_{\del}, P^{\ilp}_{\exh})$ corresponds to a feasible solution~$D^{\ilp}$ to~$\mathcal{I}$ with cardinality being the value of the ILP for this solution:
   ~$$D^{\ilp} = \bigcup_{P_i \in P^{\ilp}_{\del}} \bigcup_{j \in \{ 1, 2, \dots, x^{\ilp}_i\}} \{p^i_j\},$$
    where~$p^i_j \in P_i$ is the~$j$-th project from~$P_i$ processed by \greedyAV in an instance~$(P,V,B,p)$.

    Obviously, all the constraints in the defined ILP are linear. Each ILP has~$\Oh{nc}$ integer variables and~$\Oh{nc}$ constraints.
    Therefore, we can use the celebrated theorem of Lenstra Jr.~\cite{Lenstra1983,Kannan1987} to solve the ILP in $2^\Oh{nc\log nc} \cdot \numProjects^\Oh{1}$ time. 
    As we need to solve the ILP for each guess, the overall running time of the algorithm is $2^\Oh{nc}\cdot 2^\Oh{nc\log nc} \cdot \numProjects^\Oh{1} \in 2^\Oh{nc\log nc} \cdot \numProjects^\Oh{1}$.
\end{proofsketch}
\fi

In our next result, we show another fixed-parameter algorithm. This time, we focus on the combined parameter of the number of different costs $\numCosts$ and the number of deleted projects $\numDeleted$. The algorithm does not use any ILP formulation{} but rather a dynamic programming-like exploration of the state space{}.

\begin{theorem}[$\star$]
    \label{thm:AV:CCDC:FPT:deletions:costs}
    For any $f\in\{\greedyAV,\greedyCost\}$, both \CCDCs{$f$} and \DCDCs{$f$} are fixed-parameter tractable when parameterized by the number of different costs~$\numCosts$ and the number of deleted projects~$\numDeleted$, combined. The same holds for the combined parameter $q+\numCosts$.
    \end{theorem}
\ifshort
\begin{proofsketch}
    \newcommand{\DP}{\texttt{T}}
    We present the algorithm for \CCDCs{\greedyAV}.
    Let $(p_1,\ldots,p_m = p)$ be the order of $\projects$ in which they are considered by the \greedyAV rule, and let $(c_1,\ldots,c_\numCosts)$ be an order of our costs such that $c_i > c_{i+1}$. Before we start the algorithm, we add an auxiliary project $p_0$ of cost $\budget + 1$ and make sure that $p_0$ is assumed the first by \greedyAV rule. Observe that this project can never be funded, and therefore, its addition does not change the solution to the problem. We also add $c_0 = \budget + 1$ to the ordering of costs. Additionally, we add after $p$ a virtual project $p_{m+1}$, which serves as a `stopper'. For this special project, it holds that it is assumed as the last project (even after the distinguished project~$p$), is always not affordable regardless of the remaining budget, but its cost is the smallest among all available costs.

    Now, we define a \emph{state} of the algorithm. Formally, the state is a triple $(p_\ell,\vec{r},\beta)$, where
    \begin{itemize}
        \item $p_\ell$ is the first project of cost $\cost(p_\ell)$ which could not be afforded,%
        \item $\vec{r} = (r_1,\ldots,r_{\numCosts})$, where $r_i\in[\numDeleted]_0$, is a vector representing the number of projects of each cost $c_i$ from the set $\{p_0,\ldots,p_{\ell-1}\}$ which are part of the solution, and
        \item $\beta\in\N$ is the budget remaining before $p_\ell$ was considered%
    \end{itemize}
    The state represents that there is a set $D'\subseteq\{p_1,\ldots,p_{\ell-1}\}$, called \emph{partial solution}, that contains exactly $r_i$ projects of each cost $c_i$ so that the remaining budget available to the \greedyAV rule before it assumes $p_\ell$ is exactly $\beta$, and that $p_\ell$ is the first project of cost $\cost(p_\ell)$ which is not affordable. That is, all projects $p_{\ell'}$, $\ell' < \ell$, of cost $\cost(p_\ell)$ are either funded or deleted.
    
    The initial state is $(p_0,\vec{\mathbf{0}},\budget)$. Observe that for this state, $D' = \emptyset$ is a corresponding partial solution. There are no projects before $p_0$, so we cannot delete any project. Moreover, $p_0$ is clearly the first project of cost $c_0$ which the \greedyAV rule cannot afford, and the budget remaining before $p_0$ (and also available after $p_0$ since $p_0$ is not affordable) is considered by the \greedyAV rule, is equal to {}$\budget$.

    Now, let $s=(p_\ell,\vec{r},\beta)$ be a state with $p_\ell \not= p_{m+1}$. For each such $s$, we create at most $(r+1)^c$ next states, each computed as follows. For every vector $\vec{d} = (d_1,\ldots,d_{\numCosts})$ such that $d_i\in[\numDeleted]$ for every $i\in[c]$, we first check whether $\sum_{i=1}^{\numCosts} r_i + d_i \leq \numDeleted$. If this condition is not satisfied, we skip this particular vector $\vec{d}$ and continue with another possibility, as it would mean that a hypothetical solution would require removing more projects than is allowed by the bound~$\numDeleted$. Otherwise, we verify that the set $\{p_{\ell+1},\ldots,p_{m-1}\}$ contains at least $d_i$ projects of cost $c_i$ for every $i\in[\numCosts]$. If not, we again skip this $\vec{d}$, as it requires us to remove more projects of some cost $c_i$ than is available in the remainder of the projects. On the other hand, if there are enough projects of each cost, we remove from $\{p_{\ell+1},\ldots,p_{m-1}\}$ exactly $d_i$ leftmost projects of every cost $c_i$, $i\in[\numCosts]$. Let $D'$ denote the set of removed projects in the previous step. Observe that since $D'\subseteq \{p_{\ell+1},\ldots,p_{m-1}\}$, the distinguished project $p$ is clearly not in $D'$. Now, we simulate the \greedyAV rule for an auxiliary instance $(\{p_{\ell+1},\ldots,p\}\setminus D',\voters,\beta)$ and stop when, for the first time, the rule cannot afford some project $p_{\ell'}$ of cost $c_i < \cost(p_\ell)$. Notice that such a situation always occurs due to the virtual project $p_{m+1}$. Therefore, we add an edge to a new state $(p_{\ell'},\vec{r}+\vec{d},\beta')$, where $\beta'$ is the remaining budget just before the \greedyAV rule assumed the project $p_{\ell'}$ in our auxiliary instance.

    Once the previous procedure terminates, we can decide on the original instance. Observe that the states form a tree. So, it is enough to check that at least one leaf of this tree is a terminal state $(p_{m+1},\vec{r},\beta)$ with $\beta \geq \cost(p)$. For running time, recall that we construct at most $\numDeleted^\Oh{\numCosts}$ children in each state. It remains to show that the depth of the tree is also a function of the parameter. To see this, let $(p_0,\vec{\mathbf{0}},\budget),(p_{\ell_1},\vec{r_1},\beta_1),\ldots,(p_{\ell_j},\vec{r_j},\beta_j)$ be the longest path from the initial state to some leaf state. Recall that for every child state $(p_{\ell_{j+1}},\vec{r_{j+1}},\beta_{j+1})$ of a state $(p_{\ell_j},\vec{r_j},\beta_j)$, it holds that $\cost(p_{\ell_{j+1}})< \cost(p_{\ell_j})$. As there are $\numCosts$ different costs, the length of this path is at most $\numCosts + 1$. Consequently, the whole tree is of size $\numDeleted^\Oh{\numCosts^2}$, which, multiplied by a polynomial on the input size, is the running time of our algorithm.
\end{proofsketch}
\fi

The previous algorithm also trivially shows that the problems of interest are also in \XP when parameterized by~$\numCosts$.

\begin{corollary}\label{thm:AV:CCDC:XP:costs}
    For any $f\in\{\greedyAV,\greedyCost\}$, both \CCDCs{$f$} and \DCDCs{$f$} are in \XP when parameterized by the number of different costs~$\numCosts$.
\end{corollary}

\section{Approximation Algorithms}\label{sec:approx}

In this section, we switch our perspective to approximation algorithms. That is, we will be interested in how close a polynomial-time algorithm can get to an optimal solution of the optimization refinements of our problems.

We again start with the project deletion operation, where we first define the optimization variants of the control problems.
The optimization variant of \CCDCs{$f$}, called \probName{$f$-Opt-CCDC}, is naturally defined as a minimization problem.
For an exhaustive $f$ (e.g., a greedy PB rule),
a feasible solution of size~$m-1$ always exists: by deleting all projects except~$p$, we ensure that~$p$ is funded.
For \DCDCs{$f$}, it might be the case that removal of any subset of projects in $\projects\setminus\{p\}$, the project $p$ is always funded. Therefore, we define \probName{$f$-Opt-DCDC} so that if it is not possible to ensure the control goal by removing any subset of projects, we allow deleting~$p$. Nevertheless, this may happen only after deleting (and considering) all other projects. By this definition, we guarantee the existence of a feasible solution of size $m$ for all instances.

We start with a trivial approximation algorithm for any exhaustive PB rule (and thus for any greedy PB rule), which checks all subsets of projects of size at most~$a$, and if none of them is feasible, it returns a feasible solution of size~$\leq m$. This yields an~$m/a$-approximation running it time~$m^a \cdot \mathrm{poly}(|\mathcal{I}|)$.

\begin{observation}
    For every fixed~$a \geq 1$,
    there exists a poly\-no\-mial-time~$m/a$-approximation algorithm for both \probName{$f$-Opt-CCDC} and \probName{$f$-Opt-DCDC},
    where~$f$ is any exhaustive rule.
\end{observation}

The above trivial algorithm appears to be essentially the best possible for both rules.
In fact, in the remainder of this section, we show that the existence of an~$m^{1-\epsilon}$-approximation for these problems would imply~${\P=\NP}$.
We begin with \CCDCs{\greedyAV}.

\begin{theorem}
\label{thm:AV:CCDC:inapx}
    For every~$\epsilon > 0$,
    both \probName{\greedyAV-Opt-CCDC} and \probName{\greedyCost-Opt-CCDC} are \NPh to approximate within a factor~$m^{1-\epsilon}$,
    even if~$|\voters|=3$.
\end{theorem}
\ifshort
\begin{proofsketch}
    For exposition, we consider~$\epsilon \geq 0.5$ and give a reduction independent of the exact value of~$\epsilon$.
    This also illustrates the main idea for~$\epsilon \in (0,0.5)$. Specifically, a polynomial algorithm achieving the approximation factor $m^{1-\epsilon}$ would solve an \NPh problem \RXthreeC (\RXthreeCs). In this problem, we are given a universe $\mathcal{U}=\{u_1,\ldots,u_{3N}\}$ and a collection $\mathcal{S}=(S_1,\ldots,S_{3N})$ of $3$-sized subsets of $\mathcal{U}$ such that each $u_i\in\mathcal{U}$ is part of exactly three sets of $\mathcal{S}$, and the goal is to decide whether there exists and exact cover $X\subseteq \mathcal{S}$.%

    Given an instance~$\mathcal{I}=(U,\mathcal{S})$ of \RXthreeCs,
    we construct an instance~$\mathcal{J}$ of \probName{\greedyAV-Opt-CCDC} as follows.%
    For each set~$S_j\in\mathcal{S}$,~$S_j = \{u_{i_1},u_{i_2},u_{i_3}\}$, we create a \emph{set-project}~$p_j$ with cost~$4^{i_1} + 4^{i_2} + 4^{i_3}$. Next, we add our \emph{distinguished project}~$p$.
    Then, we define two kinds of \emph{guard-projects}.
    Namely, we define~$N^3$ \emph{expensive-guard-projects}~$h_1,\ldots,h_{N^3}$ and~$N^3$ \emph{cheap-guard-projects}~$g_1,\ldots,g_{N^3}$.
    Therefore, the total number of projects is~$m = 2N^3+3N+1$.
    The cost of the distinguished project~$p$ is~$\sum_{i=1}^{3N} 4^i$ and for every~$i\in[N^3]$,
    we set~$\cost(g_i) = \cost(p) + 1$ and~$\cost(h_i) = \cost(g_i) + \cost(p) = 2\cdot\cost(p)+1$.
    That is, the cheap-guard-projects are only one unit more expensive than~$p$, and the expensive-guard-projects cost the sum of the costs of~$p$ and a cheap-guard-project.  
    This is important to ensure that the budget left after we delete set-projects corresponding to an exact cover 
    is exactly the cost of~$p$.
    Expensive-guard-projects protect from obtaining a solution corresponding to covering elements too many times.
    
    The set of voters consists of just three voters.
    The first voter approves all projects except~$p$.
    The second voter approves the set-projects and the expensive-guard-projects.
    The third voter approves only the set-projects.
    Such a preference profile ensures that, regardless of the tie-breaking order, the method processes first all set-projects, then all expensive-guard-projects, then all cheap-guard-projects, and as the last one, project~$p$.
    To complete the construction, we set~$\budget = 3 \cdot \sum_{i=1}^{3N} 4^i= 3 \cdot \cost(p) = \sum_{j=1}^{3N} \cost(p_j)$.
    
    The following claim states the optimum solution value gap between \Yes and \No instances. For a \Yes-instance, a feasible solution of size~$N$ exists, while for a \No-instance, every feasible solution has size at least~$N^3$.

    \ifshort
    \begin{claim}[$\star$]
    \label{claim:iapx-yes-instance}
    \label{claim:iapx-no-instance}
        $\mathcal{J}$ admits a solution $S$ of size~$N$ if and only if $\mathcal{I}$ is a \Yes-instance. Otherwise, a solution $S$ has size ~$\geq N^3$.
    \end{claim}
    \fi
    
    Hence, an~$(N^2-1)$-approximation algorithm could distinguish between \Yes- and a \No-instances of RX3C in polynomial time, which is unlikely.
\end{proofsketch}
\fi

Next, we show an analogous result for \probName{$f$-Opt-DCDC}.

\ifshort
\begin{theorem}[$\star$]
\fi
\label{thm:AV:DCDC:inapx}
    For every~$\epsilon > 0$, both
    \probName{\greedyAV-Opt-DCDC} and \probName{\greedyCost-Opt-DCDC} are \NPh to approximate within a factor~$m^{1-\epsilon}$,
    even if~$|\voters|=3$.
\end{theorem}

\section{Conclusions}\label{sec:conclusions}

We initiated a study of parameterized complexity and approximability for candidate control in participatory budgeting elections. Our work provides the first systematic algorithmic perspective on this problem and opens several promising avenues for future research. 

First, an immediate question is whether parameterization by the number of projects~$c$ is tight, i.e., whether our control problems are \Wh with respect to this parameter, or admit an \FPT algorithm. Second, it would be interesting to investigate whether some results extend to other rules used in participatory budgeting, such as the Method of Equal Shares~\cite{PetersS2020} or Phragmén’s method~\cite{BrillFJL2024}---rules also considered in the work of Faliszewski \emph{et al.}~\cite{FaliszewskiJKPSSS2025a}. Next, additional types of control operations could be studied, such as adding or deleting voters. Finally, given our strong inapproximability results, it would be worthwhile to explore the interplay between parameterized and approximation approaches, as studied, e.g, by~Bredereck \emph{et al.}~\cite{BredereckCFNN16} and Faliszewski \emph{et al.}~\cite{FaliszewskiMS2021}.

\begin{credits}
    \subsubsection{\ackname} This project has received funding from the European Research Council (ERC) under the European Union’s Horizon 2020 research and innovation programme (grant agreement No 101002854) and was co-funded by the European Union under the project Robotics and advanced industrial production (reg. no. CZ.02.01.01/00/22\_008/0004590).

    \begin{center}
        \vspace{0.25cm}
    	\includegraphics[width=3.5cm]{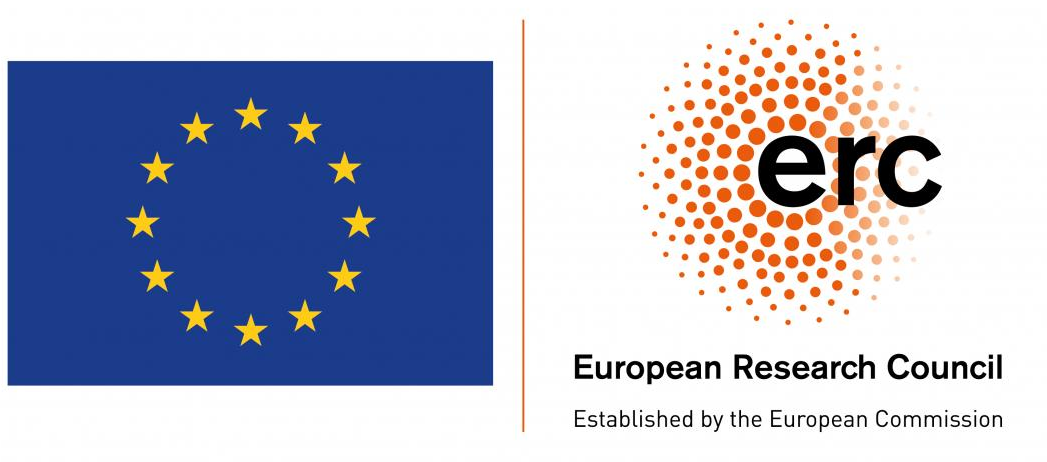}
    \end{center}

%
\end{credits}

\bibliographystyle{splncs04}
\bibliography{references}

\begin{thebibliography}{10}
\providecommand{\url}[1]{\texttt{#1}}
\providecommand{\urlprefix}{URL }
\providecommand{\doi}[1]{https://doi.org/#1}

\bibitem{AzizGGMMW2015}
Aziz, H., Gaspers, S., Gudmundsson, J., Mackenzie, S., Mattei, N., Walsh, T.:
  Computational aspects of multi-winner approval voting. In:
  \iflong{}Proceedings of the 14th International Conference on Autonomous
  Agents and Multiagent Systems, \fi{}{AAMAS}~'15. pp. 107--115. {IFAAMAS}
  (2015)

\bibitem{AzizS2021}
Aziz, H., Shah, N.: Participatory budgeting: Models and approaches. In:
  Pathways Between Social Science and Computational Social Science, pp.
  215--236. Computational Social Sciences, Springer (2021).
  \doi{10.1007/978-3-030-54936-7_10}

\bibitem{BartholdiTT1989}
{Bartholdi III}, J.J., Tovey, C.A., Trick, M.A.: The computational difficulty
  of manipulating an election. Social Choice and Welfare  \textbf{6}(3),
  227--241 (1989). \doi{10.1007/BF00295861}

\bibitem{BartholdiTT1992}
{Bartholdi III}, J.J., Tovey, C.A., Trick, M.A.: How hard is it to control an
  election? Mathematical and Computer Modelling  \textbf{16}(8-9),  27--40
  (1992). \doi{10.1016/0895-7177(92)90085-Y}

\bibitem{BaumeisterH2023}
Baumeister, D., Hogrebe, T.: On the complexity of predicting election outcomes
  and estimating their robustness. {SN} Computer Science  \textbf{4}(4), ~362
  (2023). \doi{10.1007/s42979-023-01725-0}

\bibitem{BetzlerBCN2012}
Betzler, N., Bredereck, R., Chen, J., Niedermeier, R.: Studies in computational
  aspects of voting - {A} parameterized complexity perspective. In: The
  Multivariate Algorithmic Revolution and Beyond. LNCS, vol.~7370, pp.
  318--363. Springer (2012). \doi{10.1007/978-3-642-30891-8_16}

\bibitem{BoehmerFJPPSSS2024}
Boehmer, N., Faliszewski, P., Janeczko, f., Peters, D., Pierczyński, G.,
  Schierreich, f., Skowron, P., Szufa, S.: Evaluation of project performance in
  participatory budgeting. In: \iflong{}Proceedings of the 33rd International
  Joint Conference on Artificial Intelligence, \fi{}{IJCAI}~'24. pp.
  2678--2686. ijcai.org (2024). \doi{10.24963/ijcai.2024/296}

\bibitem{BoehmerFJK2023}
Boehmer, N., Faliszewski, P., Janeczko, {\L}., Kaczmarczyk, A.: Robustness of
  participatory budgeting outcomes: Complexity and experiments. In:
  \iflong{}Proceedings of the 16th International Symposium on Algorithmic Game
  Theory, \fi{}{SAGT}~'23. LNCS, vol. 14238, pp. 161--178. Springer (2023).
  \doi{10.1007/978-3-031-43254-5_10}

\bibitem{BredereckCFNN16}
Bredereck, R., Chen, J., Faliszewski, P., Nichterlein, A., Niedermeier, R.:
  Prices matter for the parameterized complexity of shift bribery. Information
  and Computation  \textbf{251},  140--164 (2016).
  \doi{10.1016/j.ic.2016.08.003}

\bibitem{Brelsford2007}
Brelsford, E.: Approximation and Elections. Master's thesis, Rochester
  Institute of Technology (2007)

\bibitem{BrelsfordFHSS2008}
Brelsford, E., Faliszewski, P., Hemaspaandra, E., Schnoor, H., Schnoor, I.:
  Approximability of manipulating elections. In: \iflong{}Proceedings of the
  23rd {AAAI} Conference on Artificial Intelligence, \fi{}{AAAI}~'08. pp.
  44--49. {AAAI} Press (2008)

\bibitem{BrillFJL2024}
Brill, M., Freeman, R., Janson, S., Lackner, M.: Phragm{\'{e}}n's voting
  methods and justified representation. Mathematical Programming
  \textbf{203}(1),  47--76 (2024). \doi{10.1007/s10107-023-01926-8}

\bibitem{Cabannes2004}
Cabannes, Y.: Participatory budgeting: {A} significant contribution to
  participatory democracy. Environment and Urbanization  \textbf{16}(1),
  27--46 (Apr 2004). \doi{10.1177/095624780401600104}

\bibitem{CaragiannisKM10}
Caragiannis, I., Kalaitzis, D., Markakis, E.: Approximation algorithms and
  mechanism design for minimax approval voting. In: \iflong{}Proceedings of the
  24th {AAAI} Conference on Artificial Intelligence, \fi{}{AAAI}~'10. pp.
  737--742. {AAAI} Press (2010). \doi{10.1609/aaai.v24i1.7615}

\bibitem{ChenFNT2017}
Chen, J., Faliszewski, P., Niedermeier, R., Talmon, N.: Elections with few
  voters: Candidate control can be easy. Journal of Artificial Intelligence
  Research  \textbf{60},  937--1002 (2017). \doi{10.1613/jair.5515}

\bibitem{ChenKNRSS2025}
Chen, J., Kaczmarek, J., N{\"{u}}sken, P., Rothe, J., Schlotter, I., Seeger,
  T.: Control in computational social choice. In: \iflong{}Proceedings of the
  34th International Joint Conference on Artificial Intelligence,
  \fi{}{IJCAI}~'25. pp. 10391--10399. ijcai.org (2025).
  \doi{10.24963/ijcai.2025/1154}

\bibitem{ConitzerSL2007}
Conitzer, V., Sandholm, T., Lang, J.: When are elections with few candidates
  hard to manipulate? Journal of the {ACM}  \textbf{54}(3) (2007).
  \doi{10.1145/1236457.1236461}

\bibitem{ConitzerW2016}
Conitzer, V., Walsh, T.: Barriers to manipulation in voting. In: Handbook of
  Computational Social Choice, pp. 127--145. Cambridge University Press (2016)

\bibitem{CyganFKLMPPS2015}
Cygan, M., Fomin, F.V., Kowalik, L., Lokshtanov, D., Marx, D., Pilipczuk, M.,
  Pilipczuk, M., Saurabh, S.: Parameterized Algorithms. Springer, Cham (2015).
  \doi{10.1007/978-3-319-21275-3}

\bibitem{DowneyF1995}
Downey, R.G., Fellows, M.R.: Fixed-parameter tractability and completeness
  {II:} {On} completeness for {W[1]}. Theoretical Computer Science
  \textbf{141}(1{\&}2),  109--131 (1995). \doi{10.1016/0304-3975(94)00097-3}

\bibitem{DowneyF2013}
Downey, R.G., Fellows, M.R.: Fundamentals of Parameterized Complexity. Texts in
  Computer Science, Springer, London (Dec 2013)

\bibitem{ElkindF2010}
Elkind, E., Faliszewski, P.: Approximation algorithms for campaign management.
  In: \iflong{}Proceedings of the 6th International Workshop on Internet and
  Network Economics, \fi{}{WINE}~'10. LNCS, vol.~6484, pp. 473--482. Springer
  (2010). \doi{10.1007/978-3-642-17572-5_40}

\bibitem{ErdelyiNRRYZ2021}
Erd{\'{e}}lyi, G., Neveling, M., Reger, C., Rothe, J., Yang, Y., Zorn, R.:
  Towards completing the puzzle: Complexity of control by replacing, adding,
  and deleting candidates or voters. Autonomous Agents and Multiagent Systems
  \textbf{35}(2), ~41 (2021). \doi{10.1007/s10458-021-09523-9}

\bibitem{Faliszewski2008}
Faliszewski, P.: Nonuniform bribery. In: \iflong{}Proceedings of the 7th
  International Joint Conference on Autonomous Agents and Multiagent Systems,
  \fi{}{AAMAS}~'08. pp. 1569--1572. {IFAAMAS} (2008)

\bibitem{FaliszewskiFPPSSST2023}
Faliszewski, P., Flis, J., Peters, D., Pierczyński, G., Skowron, P., Stolicki,
  D., Szufa, S., Talmon, N.: Participatory budgeting: Data, tools and analysis.
  In: \iflong{}Proceedings of the 32nd International Joint Conference on
  Artificial Intelligence, \fi{}{IJCAI}~'23. pp. 2667--2674. ijcai.org (2023).
  \doi{10.24963/ijcai.2023/297}

\bibitem{FaliszewskiHH2009}
Faliszewski, P., Hemaspaandra, E., Hemaspaandra, L.A.: How hard is bribery in
  elections? Journal of Artificial Intelligence Research  \textbf{35},
  485--532 (2009). \doi{10.1613/jair.2676}

\bibitem{FaliszewskiHH2015}
Faliszewski, P., Hemaspaandra, E., Hemaspaandra, L.A.: Weighted electoral
  control. Journal of Artificial Intelligence Research  \textbf{52},  507--542
  (2015). \doi{10.1613/jair.4621}

\bibitem{FaliszewskiJKPSSS2025a}
Faliszewski, P., Janeczko, f., Knop, D., Pokorn{\'{y}}, J., Schierreich, f.,
  S{\l}uszniak, M., Sornat, K.: Participatory budgeting project strength via
  candidate control. In: \iflong{}Proceedings of the 34th International Joint
  Conference on Artificial Intelligence, \fi{}{IJCAI}~'25. pp. 3821--3829.
  ijcai.org (2025). \doi{10.24963/ijcai.2025/425}

\bibitem{FaliszewskiMS2021}
Faliszewski, P., Manurangsi, P., Sornat, K.: Approximation and hardness of
  shift-bribery. Artificial Intelligence  \textbf{298},  103520 (2021).
  \doi{10.1016/j.artint.2021.103520}

\bibitem{FaliszewskiR2016}
Faliszewski, P., Rothe, J.: Control and bribery in voting. In: Handbook of
  Computational Social Choice, pp. 146--168. Cambridge University Press (2016)

\bibitem{FaliszewskiST2017}
Faliszewski, P., Skowron, P., Talmon, N.: Bribery as a measure of candidate
  success: Complexity results for approval-based multiwinner rules. In:
  \iflong{}Proceedings of the 16th Conference on Autonomous Agents and
  MultiAgent Systems, \fi{}{AAMAS}~'17. pp. 6--14. {IFAAMAS} (2017)

\bibitem{HemaspaandraHR2007}
Hemaspaandra, E., Hemaspaandra, L.A., Rothe, J.: Anyone but him: The complexity
  of precluding an alternative. Artificial Intelligence  \textbf{171}(5-6),
  255--285 (2007). \doi{10.1016/j.artint.2007.01.005}

\bibitem{Kannan1987}
Kannan, R.: Minkowski's convex body theorem and integer programming.
  Mathematics of Operations Research  \textbf{12}(3),  415--440 (1987).
  \doi{10.1287/MOOR.12.3.415}

\bibitem{KellerHH2019}
Keller, O., Hassidim, A., Hazon, N.: Approximating weighted and priced bribery
  in scoring rules. Journal of Artificial Intelligence Research  \textbf{66},
  1057--1098 (2019). \doi{10.1613/jair.1.11538}

\bibitem{KusekBFKK23}
Kusek, B., Bredereck, R., Faliszewski, P., Kaczmarczyk, A., Knop, D.: Bribery
  can get harder in structured multiwinner approval election. In:
  \iflong{}Proceedings of the 22nd International Conference on Autonomous
  Agents and Multiagent Systems, \fi{}{AAMAS}~'23. pp. 1725--1733. {IFAAMAS}
  (2023)

\bibitem{Lenstra1983}
Lenstra~Jr., H.W.: Integer programming with a fixed number of variables.
  Mathematics of Operations Research  \textbf{8}(4),  538--548 (1983).
  \doi{10.1287/MOOR.8.4.538}

\bibitem{MeirPRZ2008}
Meir, R., Procaccia, A.D., Rosenschein, J.S., Zohar, A.: Complexity of
  strategic behavior in multi-winner elections. Journal of Artificial
  Intelligence Research  \textbf{33},  149--178 (2008). \doi{10.1613/jair.2566}

\bibitem{Niedermeier2006}
Niedermeier, R.: Invitation to Fixed-Parameter Algorithms. Oxford Lecture
  Series in Mathematics and Its Applications, Oxford University Press (2006).
  \doi{10.1093/ACPROF:OSO/9780198566076.001.0001}

\bibitem{PetersS2020}
Peters, D., Skowron, P.: Proportionality and the limits of welfarism. In:
  Bir{\'{o}}, P., Hartline, J.D., Ostrovsky, M., Procaccia, A.D. (eds.)
  \iflong{}Proceedings of the 21st {ACM} Conference on Economics and
  Computation, \fi{}{EC}~'20. pp. 793--794. {ACM} (2020).
  \doi{10.1145/3391403.3399465}

\bibitem{ReyEH2021}
Rey, S., Endriss, U., de~Haan, R.: Shortlisting rules and incentives in an
  end-to-end model for participatory budgeting. In: \iflong{}Proceedings of the
  30th International Joint Conference on Artificial Intelligence,
  \fi{}{IJCAI}~'21. pp. 370--376. ijcai.org (2021).
  \doi{10.24963/ijcai.2021/52}

\bibitem{ReySM2025}
Rey, S., Schmidt, F., Maly, J.: The (computational) social choice take on
  indivisible participatory budgeting (2025)

\bibitem{TalmonF2019}
Talmon, N., Faliszewski, P.: A framework for approval-based budgeting methods.
  In: \iflong{}Proceedings of the 33rd {AAAI} Conference on Artificial
  Intelligence, \fi{}{AAAI}~'19. pp. 2181--2188. {AAAI} Press (2019).
  \doi{10.1609/aaai.v33i01.33012181}

\bibitem{Yang2014}
Yang, Y.: Election attacks with few candidates. In: \iflong{}Proceedings of the
  21st European Conference on Artificial Intelligence, \fi{}{ECAI}~'14. FAIA,
  vol.~263, pp. 1131--1132. {IOS} Press (2014).
  \doi{10.3233/978-1-61499-419-0-1131}

\bibitem{Yang2019}
Yang, Y.: Complexity of manipulating and controlling approval-based multiwinner
  voting. In: \iflong{}Proceedings of the 28th International Joint Conference
  on Artificial Intelligence, \fi{}{IJCAI}~'19. pp. 637--643. ijcai.org (2019)

\bibitem{Yang2020}
Yang, Y.: On the complexity of destructive bribery in approval-based
  multi-winner voting. In: \iflong{}Proceedings of the 19th International
  Conference on Autonomous Agents and Multiagent Systems, \fi{}{AAMAS}~'20. pp.
  1584--1592. IFAAMAS (2020)

\bibitem{YangG2015}
Yang, Y., Guo, J.: How hard is control in multi-peaked elections: {A}
  parameterized study. In: \iflong{}Proceedings of the 14th International
  Conference on Autonomous Agents and Multiagent Systems, \fi{}{AAMAS}~'15. pp.
  1729--1730. {IFAAMAS} (2015)

\bibitem{ZuckermanPR2009}
Zuckerman, M., Procaccia, A.D., Rosenschein, J.S.: Algorithms for the
  coalitional manipulation problem. Artificial Intelligence  \textbf{173}(2),
  392--412 (2009). \doi{10.1016/j.artint.2008.11.005}

\end{thebibliography}

\end{document}